\documentclass[journal]{IEEEtran}
\usepackage{lineno,hyperref}
\usepackage{colortbl}
\usepackage{amsmath,amssymb,amsfonts}
\usepackage{algorithmic}
\usepackage{graphicx}
\usepackage{float}
\usepackage{orcidlink}
\usepackage{algorithm}
\usepackage{array}
\usepackage{url}
\usepackage{cite}
\usepackage{booktabs}
\usepackage{multirow}
\usepackage{tabularx}
\usepackage{stmaryrd}
\usepackage{varwidth}
\newcommand{\D}[1]{\mathrm{d}{#1}}
\usepackage[nolist, nohyperlinks]{acronym}
\begin{document}

%\title{EnhanceBridge: Enhancing Speech with Consistent Brownian Bridge}
\title{SE-Bridge: Speech Enhancement with Consistent Brownian Bridge}
\author{Zhibin Qiu\,{\orcidlink{0009-0004-5154-2894}},~\IEEEmembership{Student Member,~IEEE}, Mengfan Fu\,{\orcidlink{0009-0001-4676-4095}},
Fuchun Sun\,{\orcidlink{0000-0003-3546-6305}},~\IEEEmembership{Fellow,~IEEE},\\ Gulila Altenbek\,{\orcidlink{0000-0002-2077-1684}}, Hao Huang\,{\orcidlink{0000-0001-6604-0951}},~\IEEEmembership{Member,~IEEE}

\thanks{This work was supported
by Opening Project of Key Laboratory of Xinjiang, China (2020D04047),
the National Key R\&D Program of China (2020AAA0107902) and NSFC (61663044, 61761041).}
\thanks{Zhibin Qiu and Mengfan Fu are master students with the School of Information Science and Engineering at Xinjiang University and have made equal contributions to this letter. Zhibin Qiu provided the main idea and conducted speech enhancement related experiments (email: 2206411193@qq.com), while Mengfan Fu was responsible for writing the paper (email: 2574503276@qq.com).  }

\thanks{
The other authors provide guidance (email: 
fcsun@mail.tsinghua.edu.cn, gla@xju.edu.cn, huanghao@xju.edu.cn),
Huang Hao and Gulila Altenbek are with the School of Information Science and Engineering, Xinjiang University, and Fuchun Sun is with the Department of Computer Science and Technology,Tsinghua University. Hao Huang is the corresponding author.}
}
\maketitle
\begin{abstract}
We propose SE-Bridge, a novel method for speech enhancement (SE). After recently applying the diffusion models to speech enhancement, we can achieve speech enhancement by solving a stochastic differential equation (SDE). Each SDE corresponds to a probabilistic flow ordinary differential equation (PF-ODE), and the trajectory of the PF-ODE solution consists of the speech states at different moments. Our approach is based on consistency model that ensure any speech states on the same PF-ODE trajectory, correspond to the same initial state. By integrating the Brownian Bridge process, the model is able to generate high-intelligibility speech samples without adversarial training. This is the first attempt that applies the consistency models to SE task, achieving state-of-the-art results in several metrics while saving 15 x the time required for sampling compared to the diffusion-based baseline. Our experiments on multiple datasets demonstrate the effectiveness of SE-Bridge in SE. Furthermore, we show through extensive experiments on downstream tasks, including Automatic Speech Recognition (ASR) and Speaker Verification (SV), that SE-Bridge can effectively support multiple downstream tasks.
\end{abstract}

\section{Introduction}

\IEEEPARstart{S}{peech} enhancement aims to enhance the quality of speech signal by separating clean speech from noisy speech, thereby improving auditory intelligibility or facilitating downstream tasks such as ASR and SV. 
SE methods can be divided into generative enhancement methods (GEMs) and discriminative enhancement methods (DEMs). 
Because GEMs directly learn the distribution of clean speech rather than the mapping between noisy and clean speech, they have better generalize to different noise \cite{welker2022speech}. These approaches allow the models to effectively enhance noisy speech even if the dataset does not contain sufficient types of noise.  Common techniques of GEMs include those based on GANs\cite{pascual2017segan,Time-Frequency2018Soni,SERGAN,CPGAN2020Liu,fu2021metricgan+}, VAEs\cite{VAE-SEAME,VAE1-SEAME,VAE2-SEAME,VAE3-SEAME,VAE4-SEAME,VAE5-SEAME}, and Normalized Flows\cite{Flow-Based-SEAME,A2021Strauss}. 
Recently, diffusion-based methods have gained significant attention.
The SE approaches based on the diffusion models achieved the state-of-the-art performance and led to better speech intelligibility in auditory tests~\cite{richter2022speech}. However, the need for multi-step denoising of the diffusion models make them unable to meet the requirements of speech enhancement for low latency.
In this letter, we propose to apply the consistency models to SE, because the consistency models can obtain high quality sample in a single step, which can effectively meet the demand for low latency in SE. Meanwhile, we use a Brownian Bridge stochastic process to gradually change the state of the samples from the distribution of clean speech to the distribution of noisy speech. Compared with the method of applying Brownian bridge to the diffusion models by~\cite{lay2023reducing}, we discovered that the combination with the Variance Exploding Stochastic Differential Equations (VE-SDE) may be unnecessary, and the direct use of the Brownian bridge process is able to achieve similar or even better enhancement results. Also, our proposed method can effectively avoid the adjustment of hyperparameters in the diffusion coefficients of the SDE equation. The consistency models enable single-step enhancement, thus avoiding the need for the empirical settings required for the step-by-step scheduling of the sampling process used in the diffusion-based models.
Besides, SE models may exhibit over-suppression in downstream tasks~\cite{p2020Bridging}. To verify the support of our proposed method for downstream tasks, we also experimentally demonstrate the effective enhancement for the performance of downstream task models.

\section{Methods}
\begin{figure*}[t]
    \centering
    \label{fig:main}
    \includegraphics[width=\textwidth]{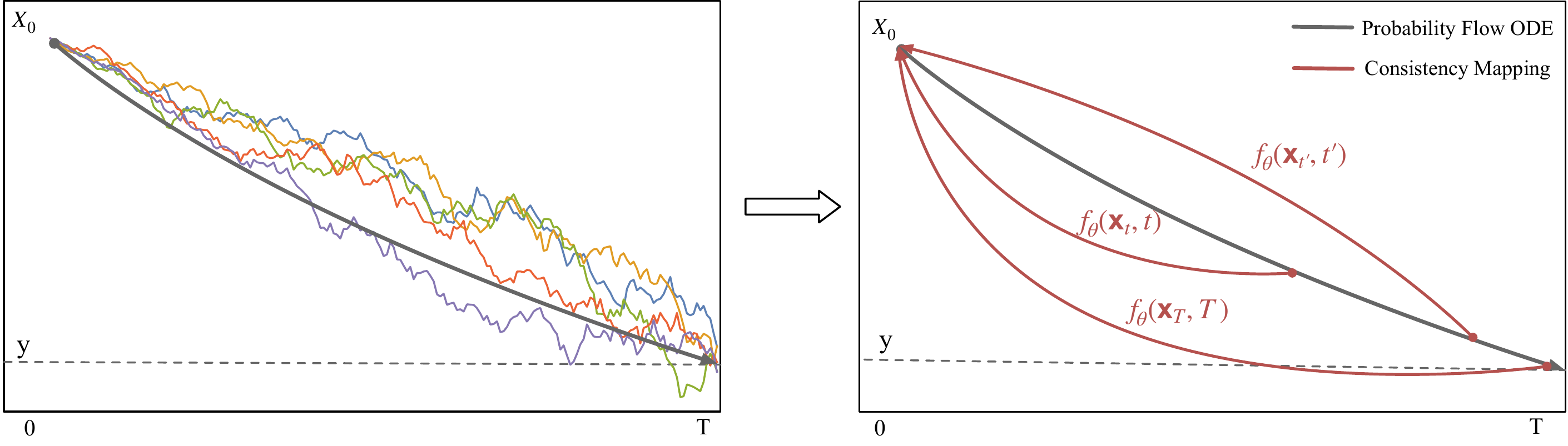}
    \caption{Overview of the proposed SE-Bridge. A thick black line in the middle of the left side diagram indicates a trajectory, of PF-ODE. By smoothly adding noise, we convert clean speech $\mathbf{x}_0$ into noisy speech $\mathbf{y}$. The remaining colored lines in the left panel represent the five simulated Brownian bridge random processes with fixed ends. On the right is a demonstration of self-consistency, where the process states on the same PF-ODE trajectory correspond to the same initial state.}
\end{figure*}
Consistency models are newly proposed generative models recently, which support sampling in one step or a few steps compared to the diffusion models. The consistency models can be trained by distillation of the pre-trained diffusion models or as stand-alone models. In this letter, we use the latter method. 

As with the diffusion-based approaches, we need to first design a stochastic diffusion process ${\{\mathbf{x}_t\}^T_{t= \epsilon}}$ modeled as a solution to a linear stochastic differential equation:
\begin{equation}
    \D{\mathbf x_t} =
            \mu(\mathbf x_t,t) 
         \D{t}+g(t)\D {\mathbf w_t},
\end{equation}
where $\mathbf{x}_t$ denotes the state of the speech sample at a certain time, called \emph{process state}, and $\mathbf{w}_t$ is the standard Wiener process.
$\mu(\cdot,\cdot):\mathbb{R}^d \rightarrow \mathbb{R}^d$ denotes the drift coefficient and $g(\cdot)$ denotes the diffusion coefficient of $\mathbf{x}_t$.
Any SDE diffusion process corresponds to a deterministic trajectory, called PF-ODE, which share the same marginal probability distribution $p(\mathbf{x})$ as the SDE,
\begin{equation}
    \D{\mathbf x} =
        \left[
            \mu(\mathbf x,  t) - \frac{1}{2} g(t)^2\nabla_{\mathbf x} \log p_t(\mathbf x)
        \right] \D{t}.
\end{equation}
Once the \emph{Score} ($\nabla_{\mathbf x} \log p_t(\mathbf x)$) is determined, we can obtain any process state on the ODE trajectory by ODE solver. 
In this letter, we design the following stochastic differential equation,
\begin{equation}
 \label{eq:3}
    \D{\mathbf{x}_t}=\frac{\mathbf{y}-\mathbf{x}_t}{1-t}\D t+\D{\mathbf{w}_t}, t\in[\epsilon,T].
\end{equation}
The stochastic process represented by this SDE satisfies the definition of a Brownian Bridge process with fixed ends, denoting the clean and the noisy speech data distribution, respectively. $\epsilon$ is a fixed small positive number to indicate the minimum processing moment and $T$ indicates the maximum processing moment.
$\mathbf{x}_\epsilon$ denotes the initial state of the speech, i.e., the clean speech, and $\mathbf{x}_T$ denotes the final state, i.e., the noisy speech $\mathbf{y}$. The left part of Fig.~\ref{fig:main} simulates five stochastic trajectories of the Brownian bridge and the corresponding PF-ODE. We can write our perturbation kernel by Eq.~\ref{eq:3}:
\begin{equation}
 \label{eq:4}
    p_{\epsilon t}(\mathbf{x}_t|\mathbf{x}_\epsilon,\mathbf{y})=\mathcal{N}_\mathbb{C}(\mathbf{x}_t;\mu(\mathbf{x}_\epsilon,\mathbf{y},t),\sigma(t)^2\emph{I}),
\end{equation}
 where $\mathcal{N}_\mathbb{C}$ denotes the circularly-symmetric complex normal distribution, while $\emph{I}$ represents the identity matrix.  Since the initial state $\mathbf{x}_0$ is known, we could use the equation (5.50, 5.53) in~\cite{s2019Applied} to get the closed-form solution of the mean and variance,
\begin{eqnarray}
    \mu(t)&=&\mathbf{x}_\epsilon(1-t)+\mathbf{y} t, \\
   \sigma(t)^2&=&t(1-t).
\end{eqnarray}
The mean moves linearly from clean to noisy speech distribution and the two ends of the variance remain zero. 
So, we eliminate the need to sample from the prior distribution (i.e., Gaussian distribution) during sampling, and can instead directly set $\mathbf{x}_T=\mathbf{y}$.
As depicted on the right side of Fig. \ref{fig:main}, we focus solely on the PF-ODE trajectory, in the right side diagram. As shown by the solid red line, our objective is to train a model $f_\theta(\mathbf{x}_t,\mathbf{y},t)$ that can map any state on the same PF-ODE trajectory, to the same initial state $\mathbf{x}_\epsilon$. This approach is aligned with the self-consistency concept proposed in consistency models~\cite{song2023consistency}.
 The training process of SE-Bridge is shown in Algorithm.\ref{alg:al}.
 The theorem 2 in~\cite{song2023consistency} states that the training of the consistency model can be independent of the pre-trained diffusion model when the time interval tends to zero, which is the theory used in this study to train the consistency model independently. Specifically, at first, sampling a pair of $ \mathbf{x, y}$ from the dataset, we can generate a pair of adjacent data points $\mathbf{x}_t^{n}, \mathbf{x}_t^{n+1}$ on the same PF-ODE trajectory efﬁciently by the perturbation kernel (Eq.\ref{eq:4}). Afterwards, we update the parameters by minimizing the output differences between the Exponential Moving Average (EMA) model and the current model. The objective function of SE-Bridge is defined as:
\begin{equation}
 \mathcal{L}^N_{SEB}(\theta,\theta^-):=\lambda(t_n)\D(f_\theta(\mathbf{x}_t^{n+1},\mathbf{y},t_{n+1}),f_{\theta^-}(\mathbf{x}_t^{n},\mathbf{y},t_n)),
\end{equation}
where $\D(\cdot)$ denotes the function that measures the distance, here $l_2$ loss is used and
$\lambda(\cdot)\in \mathbb{R}^+$ is a positive weighting function, and we set $\lambda(t_n) = 1$. We discretize the time horizon $[\epsilon,T]$ into $N-1$ sub-intervals with boundaries $t_1 = \epsilon <t_2 < ... < t_N = T$ using the following equation:
 \begin{equation}
     t_i=(\epsilon^{1/\rho}+\frac{i-1}{N-1}(T^{1/\rho}-\epsilon^{1/\rho}))^\rho.
 \end{equation}
 This equation converts discrete time steps into specific moments, and our parameter settings are consistent with~\cite{karras2022elucidating}, where $\rho=7$.
SE-Bridge needs to satisfy the self-consistency of the consistency model, and we first need to parameterize it as follows:
\begin{equation}
    f_\theta(\mathbf{x}_t,\mathbf{y},t)=c_{\text{skip}}(t)\mathbf{x}_t+c_{\text{out}}(t)F_\theta(\mathbf{x}_t,\mathbf{y},t),
\end{equation}
where $F_\theta$ is a deep neural network whose inputs and outputs have the same dimension. The settings of $c_{\text{skip}}(t)$ and $c_{\text{out}} (t)$  are to satisfy $c_{\text{skip}}(0)=1, c_{\text{out}}(0)=0$ to ensure their boundary condition $f_\theta(\mathbf{x},\mathbf{y},0) = \mathbf{x}$ as well as the requirements of differentiable.
EMA is used for training, with $f_{\theta^-}$ denoting the target network and $f_{\theta}$ denoting the online network. To improve the stability of the training process, we update ${\theta^-}$ using the following equation:
 \begin{equation}
     \theta^-\leftarrow \text{stopgrad}(\mu \theta^- + (1-\mu) \theta),
 \end{equation}
where $\mu$ depends on the current step (k)  during training and the detail in~\cite{song2023consistency}.
After the model $f_\theta$ is well trained, we input the noisy speech $\mathbf{y}$ and the maximum moment T directly into the model during the sampling process, to get the enhanced speech $\hat{\mathbf{x}}_\epsilon=f_\theta(\mathbf{y},\mathbf{y},T)$ by one step.

\begin{algorithm}[tp]
  %\footnotesize
  \caption{Training of SE-Bridge}
  \label{alg:al}
  \begin{algorithmic}
  \STATE {$\textbf{Input:}$ dataset $\mathcal{D}$, initial model parameter $\theta$, learning rate $\eta$, step schedule $N(\cdot)$, EMA decay rate schedule $\mu(\cdot), \D(\cdot, \cdot)$ and $\lambda(\cdot)$}
  \STATE {$\theta^- \leftarrow \theta$ and $k \leftarrow 0 $}
  \STATE {$\textbf{repeat}$}
  \STATE {Sample $\mathbf{x,y}\sim \mathcal{D}$, and $n\sim \mathcal{U}\llbracket{1,N(k)-1}\rrbracket$}
  \STATE {Sample $\mathbf{z}\sim \mathcal{N}(0,\emph{I})$}
  \STATE {$\mathbf{x}_t^{n} = \mathbf{y}t_n + \mathbf{x}(1-t_n) + \sqrt{t_n(1-t_n)}\mathbf{z}$}
  \STATE {$\mathbf{x}_t^{n+1} = \mathbf{y}t_{n+1}+\mathbf{x}(1-t_{n+1}) + \sqrt{t_{n+1}(1-t_{n+1})}\mathbf{z}$}
  \STATE {$\mathcal{L}(\theta,\theta^-)\leftarrow \lambda(t_n) \D(f_\theta(\mathbf{x}_t^{n+1},\mathbf{y},t_{n+1}),f_{\theta^-}(\mathbf{x}_t^{n},\mathbf{y},t_n))$}\\
  \STATE {$\theta \leftarrow \theta- \eta \nabla_\theta \mathcal{L}(\theta,\theta^-)$}
  \STATE {$\theta^- \leftarrow \text{stopgrad}(\mu(k)\theta^- +(1-\mu(k))\theta)$}
  \STATE {$k \leftarrow k+1 $}
  \STATE {$\textbf{until}$ convergence}
  \end{algorithmic}
\end{algorithm}

\section{Experiments}
In this section, we briefly introduce the dataset, evaluation metrics, and relevant configurations of the experiments. Our experiments are mainly divided into two parts, the first part is the experiment of SE, which contains the comparison of SE-Bridge with other GEMs and DEMs. The second part is the experiments for downstream tasks. 
% In addition, the source code of SE-Bridge will be found online \footnote{
% https://github.com/zhibinQiu/SE-Bridge.git.
% }
\subsection{Baselines}

To evaluate the efficacy of our approach, we conducted experimental comparisons with multiple typical speech enhancement models, including six GEMs and two DEMs. Among these models, SGMSE+~\cite{richter2022speech} served as our primary baseline. SGMSE+ builds upon SGMSE~\cite{welker2022speech} and replaces the original deep complex U-Net with NCSN++~\cite{song2020score}. SGMSE+ delivers superior performance among generative models, comparable to the state-of-the-art discriminative model performance, and the better generalization. 
We selected SGMSE+ as our baseline to showcase our proposed consistency model based speech enhancement method's superiority over the best diffusion model approach. 
\subsection{Datasets}
\label{sec:datasets}
\paragraph{SE} 
We utilized the same datasets as our baseline~\cite{richter2022speech}: WSJ0-CHiME3 and VoiceBank-DEMAND~\cite{wsj0, barker2015third, voicebank-corpus}. For WSJ0-CHiME3, we employed the noise mixture method outlined in~\cite{richter2022speech}. The VoiceBank-DEMAND dataset, a widely recognized benchmark dataset for single-channel speech enhancement. In this study, we down-sampled the original waveforms from 48 kHz to 16 kHz.

\paragraph{ASR and SV}
To evaluate our proposed method on two downstream tasks, we created a test set by adding noise to the clean speech in the Librispeech~\cite{panayotov2015librispeech} test set. We used the CHiME3 dataset to generate the noise. The process of adding noise was similar to that used in the WSJ0-CHiME3 dataset. Specifically, we randomly selected a noise from the CHiME3 noise set and uniformly sampled a signal-to-SNR from 0-20 dB. We then mixed the noise and speech to create the test set, which we call the Librispeech-CHiME3 test set.

\begin{table}[t]
    \centering
    \caption{Experimental results of speech enhancement under matched condition. Values before "/" indicate results on WSJ0-CHiME3, and values after "/" indicate results on VoiceBank-DEMAND. Among them, STCN~\cite{richter2020speech} and RVAE~\cite{bie2021unsupervised} use WSJ0~\cite{wsj0} and VoiceBank~\cite{voicebank-corpus} datasets for unsupervised training.
}
\begin{tabular}{c|c|ccc}
\toprule 
Method & Type & PESQ($\uparrow$) & ESTOI($\uparrow$) & SI-SDR[dB]($\uparrow$) \\
\midrule
\midrule
Noisy   & -  &  $1.70/1.97 $ & $0.78/0.79 $ & $10.0/8.4 $ \\
\midrule
STCN~\cite{richter2020speech}  &G  & $2.01/1.80 $ & $0.81/0.79 $ & $13.5/11.9 $ \\
RVAE~\cite{bie2021unsupervised} &G  & $2.31/2.08 $ & $0.85/0.82 $ & $15.8/13.9 $ \\
SE-Flow~\cite{bie2021unsupervised} &G  & $2.30/2.28 $ & $0.76/0.77 $ & $15.6/16.3 $ \\ 
CDiffuSE~\cite{CD-SEAME} &G  & $2.15/2.46 $ & $0.80/0.79 $ & $7.3/12.6 $  \\ 
SGMSE~\cite{welker2022speech}  &G  & $2.28/2.28 $ & $0.86/0.80 $ & $14.8/16.2 $ \\ 
SGMSE+~\cite{richter2022speech} &G  & $2.96/2.93 $ & $0.92/\mathbf{0.87} $ & $18.3/17.3 $ \\
SE-Bridge &G  & $2.99/2.97 $        &  $\mathbf{0.95}/\mathbf{0.87}$   &$ \mathbf{20.3}/\mathbf{19.9}$ \\
\midrule
MetricGAN+~\cite{fu2021metricgan+}  &D  & $\mathbf{3.03}/\mathbf{3.13} $ & $0.88/0.83 $ & $10.5/8.5$ \\
Conv-TasNet~\cite{luo2019conv}  &D & $2.99/2.84 $ & $ 0.93/0.85 $ & $ 19.9/19.1 $ \\
\midrule
\bottomrule
\end{tabular}
    \label{tab:results:matched}
\end{table}

\begin{table}[tp]
    \centering
    \caption{Speech enhancement results under mismatched condition, the numbers in parentheses indicate the relative change in performance under mismatched condition and matched condition.}
    \begin{tabular}{c|ccc}
        \toprule 
        Method  & PESQ ($\uparrow$)& ESTOI ($\uparrow$)  & SI-SDR[dB] ($\uparrow$)  \\
        \midrule
        \midrule
        Noisy & 1.70  & 0.78 & 10.0  \\
        \midrule
        SGMSE+~\cite{richter2022speech}  & $2.48(-15.3\%)$ & $0.90(3.4\%)$  & $16.2(-6\%)$  \\
        SE-Bridge   & $\mathbf{2.51(-15.5\%)}$ & $\mathbf{0.91(4.6\%)}$ & $\mathbf{18.6(-6.5\%)}$ \\
        \midrule
        % UMX$^*$~\cite{uhlich2020open}  & D  & 2.35  & 0.83 & 14.0 \\
        Conv-TasNet~\cite{luo2019conv}    & $2.40(-15.5\%)$ &  $0.88(3.5\%)$ & $15.2(-20.4\%)$ \\
        MetricGAN+~\cite{fu2021metricgan+}  & 2.13(-32\%)  & 0.76(-8.4\%) &  6.8(-20\%) \\
        \bottomrule
    \end{tabular}
    \label{tab:results:mismathced}
\end{table}
\subsection{Performance metrics}
We used the most commonly used evaluation metrics on three tasks.
\paragraph{SE} 
We use Perceptual Evaluation of Speech Quality (PESQ) as our SE metric, with scores ranging from 1 to 4.5 for PESQ, one of the most common metrics for single-channel speech enhancement tasks.  In addition to this, we also use Extended Short-Time Objective Intelligibility (ESTOI, the scores range from 0 to 1),
Scale Invariant Signal to Distortion Ratio (SI-SDR), measured in dB, the larger the value means the better the performance of enhancement.
\paragraph{ASR}  we report Word Error Rate (WER), which is a standard metric in ASR that reflects the word-level edit distance between ASR model output and the ground-truth transcription, the lower its value, the better the recognition performance.
\paragraph{SV} we report Equal Error Rate (EER), whose purpose is to detect whether two different spoken utterances are spoken by the same person. Lower values indicate better performance.
\subsection{Configuration}
\paragraph{SE}
The data input representation and the DNN network architecture we use are consistent with~\cite{richter2022speech}, who makes the Conditional Score Network (NCSN++) structure support complex spectrograms. Specifically, we use the real and imaginary parts of the input data into the network after combining them in the channel dimension.
We trained the deep neural network on a Nvidia 3090, using the Adam optimizer with an initial learning rate of $10^{-4}$ and the batch size set to $8$. In addition, we used a gradient accumulation strategy, with gradient updates every four batches. 
The maximum number of discrete steps N is set to $30$, and we set the minimum process time $\epsilon = 0.001$ maximum process moment $T = 0.999$ to ensure numerically stable.  
\paragraph{ASR and SV}
Two GEMs and two DEMs are used for the downstream tasks of ASR and SV, respectively. In terms of dataset selection, the training data for SE-Bridge and SGMSE+ is WSJ0-CHiME3, and the training data for MetricGAN+ and Conv-TasNet was VoiceBank-DEMAND. Their test sets were both Librispeech-CHiME3. We retrained these four models to compare the performance of the test set in two downstream tasks before and after speech enhancement, respectively.
For ASR and SV, we use pre-trained models from the SpeechBrain~\cite{ravanelli2021speechbrain} toolkit. 
For ASR model, we use Transformer\cite{Attention2017Ashish} + TransformerLM model pre-trained on LibriSpeech~\cite{panayotov2015librispeech}. 
For SV model, we use ECAPA-TDNN embedding model~\cite{Desplanques2020}, pre-trained on VoxCeleb~\cite{Nagrani2017}.

\section{Results}
\subsection{SE}
\paragraph{Results on matched condition}
SE-Bridge and other recent GEMs were trained on both the WSJ0-CHIME3 and VoiceBank-DEMAND datasets and tested under matched conditions (training and testing on the same dataset). Table~\ref{tab:results:matched} shows that SE-Bridge achieved the best results in terms of ESTOI and SI-SDR on both datasets. It is worth noting that our model's SI-SDR showed a significant improvement over other models on both datasets. In terms of PESQ, the discriminative model achieved the best results, with MetricGAN+ slightly outperforming SE-Bridge. However, MetricGAN+ performed poorly in MUSHRA~\cite{series2014method} listening experiments, and subsequent experiments revealed that it had worse generalization than SE-Bridge.

\paragraph{Results on mismatched condition}
The mismatch between the test and training sets can, to some extent, reflect the generalization performance of the model. Table~\ref{tab:results:mismathced} reports the results of the  mismatch condition. The PESQ and SI-SDR performance of both the GEMs and DEMs decreased, with the DEMs decreasing significantly, especially in the SI-SDR, where the GEMs clearly dominated. This is related to the property of generative models to learn data distribution directly. Our model achieves the best performance on three metrics and can be considered to have learned the prior distribution of clean speech.
\paragraph{RTF}
SGMSE+~\cite{richter2022speech} has the same network size as SE-Bridge and it is considered as state-of-the-art diffusion-based speech enhancement method. 
We calculate the real-time factor (RTF) of SGMSE+~\cite{richter2022speech} using its sampling configuration when it achieves best performance.
Then, compared to the RTF when SE-Bridge achieves optimal performance, our model has a 15-fold \footnote{Average processing time for all the test audio ﬁles on an NVIDIA GeForce RTX 3090 GPU, in a machine with an Intel Xeon Gold 5218 CPU @ 2.30GHz.} reduction in RTF while the SE performance is superior.
\subsection{ASR and SV}
The downstream tasks usually reflect whether the speech enhancement models can retain sufficient speech information while improving speech intelligibility. Table \ref{tab:results:down_task} reports the experimental results of the downstream tasks. All four models effectively support the two downstream tasks, with our proposed model exceeding the performance of the others. There is a correlation between the performance of the downstream tasks and the PESQ, as in the mismatch condition, so the performance of both discriminative models is also relatively poor, which can also reflect the advantage of the generative speech enhancement approaches in the downstream tasks.
\begin{table}[tp]
\centering
    \caption{Experimental results of downstream tasks using the SE-Bridge enhanced Librispeech-CHiME3 test set.}
\begin{tabular}{c|c|c|c}
\toprule 
  & $\emph{SE}$& {$\emph{ASR}$} & {$\emph{SV}$}    \\
   & PESQ $\uparrow$ & WER $($\%$)\downarrow$ & EER $($\%$)\downarrow$  \\
\toprule
Clean (Upper bound) & 4.64 & 2.52 & 1.42 \\
\toprule
    Noisy & 1.63 & 14.96 & 3.54\\
    SGMSE+~\cite{welker2022speech}  &2.45  & 10.34 & 2.01   \\
    SE-Bridge (Proposed)  & $\mathbf{2.55}$  & $\mathbf{9.62}$ & $\mathbf{1.87}$  \\
    \midrule
    Conv-TasNet~\cite{luo2019conv}   & 2.48  & 11.33 & 2.36 \\
    MetricGAN+~\cite{fu2021metricgan+}   & 2.09  & 10.67 & 2.19 \\
    
\midrule
\bottomrule
\end{tabular}
    \label{tab:results:down_task}
\end{table}
\section{Conclusion}
In this letter, we propose a novel speech enhancement paradigm using the consistency models with the Brownian Bridge stochastic process. 
Our method is simple, yet highly effective. It not only improves the intelligibility of the noisy speech, but also effectively reduces the RTF when compared to diffusion-based approaches.
Our experimental results show that our proposed method outperforms the baselines, resulting in better speech intelligibility and better performance in downstream tasks.
Furthermore, our proposed SE-Bridge method can potentially be applied to other tasks such as phase recovery and image denoising. Overall, our proposed method provides a new direction for speech enhancement research and opens up opportunities for further exploration in this area.

\bibliographystyle{IEEEtran}
\bibliography{main}

\begin{thebibliography}{10}
\providecommand{\url}[1]{#1}
\csname url@samestyle\endcsname
\providecommand{\newblock}{\relax}
\providecommand{\bibinfo}[2]{#2}
\providecommand{\BIBentrySTDinterwordspacing}{\spaceskip=0pt\relax}
\providecommand{\BIBentryALTinterwordstretchfactor}{4}
\providecommand{\BIBentryALTinterwordspacing}{\spaceskip=\fontdimen2\font plus
\BIBentryALTinterwordstretchfactor\fontdimen3\font minus
  \fontdimen4\font\relax}
\providecommand{\BIBforeignlanguage}[2]{{%
\expandafter\ifx\csname l@#1\endcsname\relax
\typeout{** WARNING: IEEEtran.bst: No hyphenation pattern has been}%
\typeout{** loaded for the language `#1'. Using the pattern for}%
\typeout{** the default language instead.}%
\else
\language=\csname l@#1\endcsname
\fi
#2}}
\providecommand{\BIBdecl}{\relax}
\BIBdecl

\bibitem{welker2022speech}
S.~Welker, J.~Richter, and T.~Gerkmann, ``Speech enhancement with score-based
  generative models in the complex {STFT} domain,'' in \emph{ISCA Interspeech},
  2022.

\bibitem{pascual2017segan}
\BIBentryALTinterwordspacing
S.~Pascual, A.~Bonafonte, and J.~Serr{\`{a}}, ``{SEGAN:} speech enhancement
  generative adversarial network,'' in \emph{Interspeech 2017, 18th Annual
  Conference of the International Speech Communication Association, Stockholm,
  Sweden, August 20-24, 2017}, F.~Lacerda, Ed.\hskip 1em plus 0.5em minus
  0.4em\relax {ISCA}, 2017, pp. 3642--3646. [Online]. Available:
  \url{http://www.isca-speech.org/archive/Interspeech\_2017/abstracts/1428.html}
\BIBentrySTDinterwordspacing

\bibitem{Time-Frequency2018Soni}
M.~H. Soni, N.~Shah, and H.~A. Patil, ``Time-frequency masking-based speech
  enhancement using generative adversarial network,'' in \emph{IEEE Int. Conf.
  on Acoustics, Speech and Signal Proc. (ICASSP)}, 2018, pp. 5039--5043.

\bibitem{SERGAN}
D.~Baby and S.~Verhulst, ``{SERGAN}: Speech enhancement using relativistic
  generative adversarial networks with gradient penalty,'' in \emph{IEEE Int.
  Conf. on Acoustics, Speech and Signal Proc. (ICASSP)}, 2019, pp. 106--110.

\bibitem{CPGAN2020Liu}
G.~Liu, K.~Gong, X.~Liang, and Z.~Chen, ``{CPGAN}: Context pyramid generative
  adversarial network for speech enhancement,'' in \emph{IEEE Int. Conf. on
  Acoustics, Speech and Signal Proc. (ICASSP)}, 2020, pp. 6624--6628.

\bibitem{fu2021metricgan+}
\BIBentryALTinterwordspacing
S.~Fu, C.~Yu, T.~Hsieh, P.~Plantinga, M.~Ravanelli, X.~Lu, and Y.~Tsao,
  ``{MetricGAN+:} an improved version of metricgan for speech enhancement,'' in
  \emph{Interspeech 2021, 22nd Annual Conference of the International Speech
  Communication Association, Brno, Czechia, 30 August - 3 September 2021},
  H.~Hermansky, H.~Cernock{\'{y}}, L.~Burget, L.~Lamel, O.~Scharenborg, and
  P.~Motl{\'{\i}}cek, Eds.\hskip 1em plus 0.5em minus 0.4em\relax {ISCA}, 2021,
  pp. 201--205. [Online]. Available:
  \url{https://doi.org/10.21437/Interspeech.2021-599}
\BIBentrySTDinterwordspacing

\bibitem{VAE-SEAME}
S.~Leglaive, L.~Girin, and R.~Horaud, ``A variance modeling framework based on
  variational autoencoders for speech enhancement,'' in \emph{IEEE Int.
  Workshop on Machine Learning for Signal Proc. (MLSP)}, 2018, pp. 1--6.

\bibitem{VAE1-SEAME}
Y.~Bando, M.~Mimura, K.~Itoyama, K.~Yoshii, and T.~Kawahara, ``Statistical
  speech enhancement based on probabilistic integration of variational
  autoencoder and non-negative matrix factorization,'' in \emph{IEEE Int. Conf.
  on Acoustics, Speech and Signal Proc. (ICASSP)}, 2018, pp. 716--720.

\bibitem{VAE2-SEAME}
S.~Leglaive, L.~Girin, and R.~Horaud, ``Semi-supervised multichannel speech
  enhancement with variational autoencoders and non-negative matrix
  factorization,'' in \emph{IEEE Int. Conf. on Acoustics, Speech and Signal
  Proc. (ICASSP)}, 2019, pp. 101--105.

\bibitem{VAE3-SEAME}
S.~Leglaive, U.~Şimşekli, A.~Liutkus, L.~Girin, and R.~Horaud, ``Speech
  enhancement with variational autoencoders and alpha-stable distributions,''
  in \emph{IEEE Int. Conf. on Acoustics, Speech and Signal Proc. (ICASSP)},
  2019, pp. 541--545.

\bibitem{VAE4-SEAME}
S.~Leglaive, X.~Alameda-Pineda, L.~Girin, and R.~Horaud, ``A recurrent
  variational autoencoder for speech enhancement,'' in \emph{IEEE Int. Conf. on
  Acoustics, Speech and Signal Proc. (ICASSP)}, 2020, pp. 371--375.

\bibitem{VAE5-SEAME}
M.~Sadeghi and X.~Alameda-Pineda, ``Mixture of inference networks for vae-based
  audio-visual speech enhancement,'' \emph{IEEE Transactions on Signal
  Processing}, vol.~69, pp. 1899--1909, 2021.

\bibitem{Flow-Based-SEAME}
A.~A. Nugraha, K.~Sekiguchi, and K.~Yoshii, ``A flow-based deep latent variable
  model for speech spectrogram modeling and enhancement,'' \emph{IEEE/ACM
  Trans. Audio, speech, Lang. Process.}, vol.~28, pp. 1104--1117, 2020.

\bibitem{A2021Strauss}
M.~Strauss and B.~Edler, ``A flow-based neural network for time domain speech
  enhancement,'' in \emph{IEEE Int. Conf. on Acoustics, Speech and Signal Proc.
  (ICASSP)}, 2021, pp. 5754--5758.

\bibitem{richter2022speech}
J.~Richter, S.~Welker, J.-M. Lemercier, B.~Lay, and T.~Gerkmann, ``Speech
  enhancement and dereverberation with diffusion-based generative models,''
  \emph{arXiv preprint arXiv:2208.05830}, 2022.

\bibitem{lay2023reducing}
B.~Lay, S.~Welker, J.~Richter, and T.~Gerkmann, ``Reducing the prior mismatch
  of stochastic differential equations for diffusion-based speech
  enhancement,'' \emph{arXiv preprint arXiv:2302.14748}, 2023.

\bibitem{p2020Bridging}
P.~Wang, K.~Tan, and D.~L. Wang, ``Bridging the gap between monaural speech
  enhancement and recognition with distortion-independent acoustic modeling,''
  in \emph{IEEE Trans. on Audio, Speech, and Language Proc. (TASLP)}, vol.~28,
  2020, pp. 39--48.

\bibitem{s2019Applied}
S.~S¨arkk¨a and A.~Solin, ``Applied stochastic differential equations,'' in
  \emph{Cambridge University Press}, 2019.

\bibitem{song2023consistency}
Y.~Song, P.~Dhariwal, M.~Chen, and I.~Sutskever, ``Consistency models,''
  \emph{arXiv preprint arXiv:2303.01469}, 2023.

\bibitem{karras2022elucidating}
\BIBentryALTinterwordspacing
T.~Karras, M.~Aittala, T.~Aila, and S.~Laine, ``Elucidating the design space of
  diffusion-based generative models,'' \emph{CoRR}, vol. abs/2206.00364, 2022.
  [Online]. Available: \url{https://doi.org/10.48550/arXiv.2206.00364}
\BIBentrySTDinterwordspacing

\bibitem{song2020score}
\BIBentryALTinterwordspacing
Y.~Song, J.~Sohl{-}Dickstein, D.~P. Kingma, A.~Kumar, S.~Ermon, and B.~Poole,
  ``Score-based generative modeling through stochastic differential
  equations,'' in \emph{9th International Conference on Learning
  Representations, {ICLR} 2021, Virtual Event, Austria, May 3-7, 2021}.\hskip
  1em plus 0.5em minus 0.4em\relax OpenReview.net, 2021. [Online]. Available:
  \url{https://openreview.net/forum?id=PxTIG12RRHS}
\BIBentrySTDinterwordspacing

\bibitem{wsj0}
J.~Garofolo, D.~Graff, D.~Paul, and D.~Pallett, ``Csr-i (wsj0) complete
  ldc93s6a,'' in \emph{Web Download. Philadelphia: Linguistic Data Consortium},
  vol.~83, 1993.

\bibitem{barker2015third}
J.~Barker, R.~Marxer, E.~Vincent, and S.~Watanabe, ``The third
  ‘chime’speech separation and recognition challenge: Dataset, task and
  baselines,'' in \emph{IEEE Workshop on Automatic Speech Recognition and
  Understanding (ASRU)}.\hskip 1em plus 0.5em minus 0.4em\relax IEEE, 2015, pp.
  504--511.

\bibitem{voicebank-corpus}
\BIBentryALTinterwordspacing
C.~Veaux, J.~Yamagishi, and S.~King, ``The voice bank corpus: Design,
  collection and data analysis of a large regional accent speech database,'' in
  \emph{2013 International Conference Oriental {COCOSDA} held jointly with 2013
  Conference on Asian Spoken Language Research and Evaluation
  (O-COCOSDA/CASLRE), Gurgaon, India, November 25-27, 2013}.\hskip 1em plus
  0.5em minus 0.4em\relax {IEEE}, 2013, pp. 1--4. [Online]. Available:
  \url{https://doi.org/10.1109/ICSDA.2013.6709856}
\BIBentrySTDinterwordspacing

\bibitem{panayotov2015librispeech}
V.~Panayotov, G.~Chen, D.~Povey, and S.~Khudanpur, ``{Librispeech}: An asr
  corpus based on public domain audio books,'' in \emph{IEEE Int. Conf. on
  Acoustics, Speech and Signal Proc. (ICASSP)}.\hskip 1em plus 0.5em minus
  0.4em\relax IEEE, 2015, pp. 5206--5210.

\bibitem{richter2020speech}
J.~Richter, G.~Carbajal, and T.~Gerkmann, ``Speech enhancement with stochastic
  temporal convolutional networks,'' \emph{ISCA Interspeech}, pp. 4516--4520,
  2020.

\bibitem{bie2021unsupervised}
X.~Bie, S.~Leglaive, X.~Alameda-Pineda, and L.~Girin, ``Unsupervised speech
  enhancement using dynamical variational autoencoders,'' \emph{IEEE/ACM
  Transactions on Audio, Speech, and Language Processing}, vol.~30, pp.
  2993--3007, 2022.

\bibitem{CD-SEAME}
Y.-J. Lu, Z.-Q. Wang, S.~Watanabe, A.~Richard, C.~Yu, and Y.~Tsao,
  ``Conditional diffusion probabilistic model for speech enhancement,'' in
  \emph{IEEE Int. Conf. on Acoustics, Speech and Signal Proc. (ICASSP)}, 2022.

\bibitem{luo2019conv}
Y.~Luo and N.~Mesgarani, ``Conv-{TasNet}: Surpassing ideal time--frequency
  magnitude masking for speech separation,'' \emph{IEEE/ACM Trans. Audio,
  speech, Lang. Process.}, vol.~27, no.~8, pp. 1256--1266, 2019.

\bibitem{ravanelli2021speechbrain}
\BIBentryALTinterwordspacing
M.~Ravanelli, T.~Parcollet, P.~Plantinga, A.~Rouhe, S.~Cornell, L.~Lugosch,
  C.~Subakan, N.~Dawalatabad, A.~Heba, J.~Zhong, J.~Chou, S.~Yeh, S.~Fu,
  C.~Liao, E.~Rastorgueva, F.~Grondin, W.~Aris, H.~Na, Y.~Gao, R.~D. Mori, and
  Y.~Bengio, ``Speechbrain: {A} general-purpose speech toolkit,'' \emph{CoRR},
  vol. abs/2106.04624, 2021. [Online]. Available:
  \url{https://arxiv.org/abs/2106.04624}
\BIBentrySTDinterwordspacing

\bibitem{Attention2017Ashish}
A.~Vaswani, N.~Shazeer, N.~Parmar, J.~Uszkoreit, L.~Jones, A.~N. Gomez, Łukasz
  Kaiser, and I.~Polosukhin, ``Attention is all you need,'' in \emph{Advances
  in Neural Inf. Proc. Systems (NeurIPS)}, 2017.

\bibitem{Desplanques2020}
\BIBentryALTinterwordspacing
B.~Desplanques, J.~Thienpondt, and K.~Demuynck, ``{ECAPA-TDNN: Emphasized
  Channel Attention, Propagation and Aggregation in TDNN Based Speaker
  Verification},'' in \emph{Proc. Interspeech 2020}, 2020, pp. 3830--3834.
  [Online]. Available: \url{http://dx.doi.org/10.21437/Interspeech.2020-2650}
\BIBentrySTDinterwordspacing

\bibitem{Nagrani2017}
\BIBentryALTinterwordspacing
A.~Nagrani, J.~S. Chung, and A.~Zisserman, ``Voxceleb: A large-scale speaker
  identification dataset,'' in \emph{Proc. Interspeech 2017}, 2017, pp.
  2616--2620. [Online]. Available:
  \url{http://dx.doi.org/10.21437/Interspeech.2017-950}
\BIBentrySTDinterwordspacing

\bibitem{series2014method}
B.~Series, ``Method for the subjective assessment of intermediate quality level
  of audio systems,'' \emph{International Telecommunication Union
  Radiocommunication Assembly}, 2014.

\end{thebibliography}
\end{document}